\documentclass[12pt]{iopart}
\usepackage{graphicx}
\usepackage{cite}
\begin{document}

\title[Fluctuation-Phase Relation between Positive and Negative Ions ...]
{Fluctuation-Phase Relation between Positive and Negative Ions on Pair-Plasma Electrostatic Waves} 

\author{W. Oohara, D. Date, and R. Hatakeyama}
\address{Department of Electronic Engineering, Tohoku University Sendai 980-8579, Japan}

\date{\today}

\begin{abstract}
Three kinds of electrostatic modes are experimentally observed to propagate 
along magnetic-field lines for the first time in the pair-ion plasma 
consisting of only positive and negative fullerene ions with an equal mass.
It is found that phase lags between the density fluctuations of 
positive and negative ions vary from 0 to $\pi$ depending on the frequency 
and is fixed at $\pi$ in the cases of ion acoustic and ion plasma waves, 
respectively.
In addition, a new mode with the phase lag about $\pi$ appears 
in an intermediate-frequency band between the acoustic and plasma waves.
\end{abstract}


\maketitle

\section{Introduction}

Plasmas comprising assemblies of charged particles, which interact 
with each other through the Coulomb force, are known to reveal various 
kinds of unique properties.
A typical plasma consists of electrons and positive ions, 
and the mass difference between negative- and positive-charged particles 
essentially causes temporal and spatial varieties of collective plasma 
phenomena. 
As one of the phenomena, linear and nonlinear properties of 
electrostatic and electromagnetic waves have been investigated in plasmas 
so far. 
Pair plasmas consisting of only positive- and negative-charged particles 
with an equal mass have been investigated experimentally 
\cite{Gibson,Surko1,Surko2,Greaves,Boehmer,Liang,Amoretti} and 
theoretically \cite{Stewart1,Iwamoto,Abdul,Stewart2,Zhao,Zank,Dubin}, 
since the pair plasmas, such as electron-positron plasma, are thought to be 
generated naturally under certain astrophysical conditions.
Both the relativistic and non-relativistic pair-plasmas are gradually 
disclosed to represent a new state of matter with unique thermodynamic 
property drastically different from ordinary electron-ion plasmas.
A comprehensive analysis of the elementary properties of the pair plasmas, 
linear- and nonlinear-collective plasma modes, has theoretically been 
developed and the experimental identification is at present desired 
to be performed.
However,the identification of the collective modes is very difficult 
in the electron-positron plasmas because the annihilation time is short
compared with the plasma period.
Therefore our attention is concentrated on the stable generation of 
a pair-ion plasma consisting of positive and negative ions 
with an equal mass and the resultant collective-mode identification.

According to our previous work on the generation of an alkali-fullerene 
plasma (K$^+$, e$^-$, C$_{60}^-$) by introducing fullerenes 
into a potassium plasma \cite{Schermann,NSato,Oohara1,Oohara2}, 
fullerenes are adopted as a candidate for the ion source in order to 
realize the pair-ion plasma, 
based on the fact that the interaction of electrons with the fullerenes 
leads to the production of both negative \cite{Jaffke,Huang} and positive 
\cite{Volpel,Matt,Vostrikov} ions.
We have developed a novel method for generating the pair-ion plasma 
which consists of only positive and negative ions with an equal mass 
using fullerenes \cite{Oohara3}, and discussed basic characteristics of 
the pair-ion plasma in terms of the differences from ordinary 
electron-ion plasmas.
Here the pair-ion plasma source is drastically improved in order to increase 
the plasma density and excite effectively the collective modes.
We mainly present properties of the electrostatic modes propagating 
along magnetic-field lines.

\section{Pair-Ion Plasma Generation}

\hspace*{\parindent}
The improved pair-ion source with ion species of fullerenes is installed 
in a grounded vacuum-chamber of 15.7 cm in diameter and 260 cm in length, 
as shown schematically in Fig.~1 \cite{Oohara3}.
A uniform magnetic field of $B$ = 0.3 T is applied by solenoid coils 
and the background gas pressure is 2$\times$10$^{-4}$ Pa.
A grounded copper cylinder (8 cm in diameter and 30 cm in length) 
with a copper annulus (3 cm in inner diameter, 8 cm in outer diameter, 
and 0.1 cm in thickness) is fixed inside a cylindrical ceramic furnace 
and heated to 500 $^\circ$C.
An electron-beam gun is set inside the copper cylinder, which consists of 
four tungsten wires connected electrically in parallel.
This wire cathode heated over 2000 $^\circ$C by resistive heating is biased 
at a voltage $V_k$ ($<$ 0 V) with respect to a grounded grid set 
at less than 0.5 cm in front of the cathode.
A stainless-steel disk of 4 cm in diameter is concentrically welded onto 
the gridded anode.
Thermionically emitted low-energy electrons ($\sim$ 0.2 eV) are accelerated 
by an electric field between the cathode and anode, forming a hollow 
electron beam.
The beam flows along magnetic ($B$) field lines and is terminated 
at the grounded annulus.
The beam energy $E_e$ can be controlled in the range of 0$-$150 eV 
by changing $V_k$.
The cylinder and the ceramic furnace have a hole (3 cm in diameter) 
on the sidewall and an oven for fullerene sublimation is set there, 
where a fullerene sample (C$_{60}$ powder) is heated at a temperature 
between 400 and 600 $^\circ$C.
The fullerene vapor produced as a result of sublimation is effused 
through a 0.3-cm-diam hole under molecular flow conditions, 
filling the cylinder.

For analytic convenience, the whole space of this plasma is divided 
into three regions (I), (II), and (III), as shown in Fig.~1.
The electron-beam region is called Region (I) as a fullerene-ion production 
region.
Positive ions C$_{60}^+$ are produced by the electron-impact ionization and 
low-energy electrons are simultaneously produced in connection 
with this process. 
Negative ions C$_{60}^-$ are produced by the attachment of 
these low-energy electrons.
The attachment occurs over a very wide energy range, extending 
to about 12 eV, and which is deserving special mention.
The gyroradius ratio between C$_{60}^+$ and electron 
$\rho_{{\rm C}_{60}^+}/\rho_{{\rm e}^-}$ is especially high ($\simeq$ 1100), 
and a preferential ambipolar-diffusion of C$_{60}^+$ and C$_{60}^-$ 
can take place in the radial ($r$) direction 
across the $B$-field lines, i.e., electrons are separated 
by a magnetic-filtering effect \cite{Sheehan}.
Only C$_{60}^+$ and C$_{60}^-$ are expected to exist in the midmost of 
the cylinder, Region (II), and the electron-free pair-ion plasma generation 
is attained here. 
C$_{60}^+$ and C$_{60}^-$ flow along the $B$-field lines and pass 
through the annular hole toward an experimental region, Region (III).
The thick copper annulus (3 cm in inner diameter, 8 cm in outer diameter, 
and 3 cm in thickness) is set between Region (II) and (III), and 
independently biased at a dc voltage $V_{an}$ and an ac voltage $V_{exc}$.
The plasma density in Region (III) can be controlled by changing $V_{an}$ 
and $V_{exc}$.
The exit position of the thick annulus is defined as $z$ = 0 cm, 
and the pair-ion plasma is terminated at a floating endplate ($z$ = 90 cm).
Plasma parameters in Region (III) are measured by Langmuir probes, 
collectors of which are prevented from being contaminated by C$_{60}$.

The generation property of the pair-ion plasma depending on 
the electron-beam energy $E_e$ is measured at $r$ = 0 cm and $z$ = 5 cm 
in Region (III) for $V_{an}$ = $V_{exc}$ = 0 V, as shown in Fig.~2.
$I_+$ and $I_-$ are the Langmuir-probe saturation currents of C$_{60}^+$ 
and C$_{60}^-$, respectively, which are considered to be in proportion to 
the plasma density.
When $E_e$ increases from 0 eV, the pair-ion plasma begins to be generated.
The plasma density once saturates, a little bit decreases around 9 eV, 
increases around 15 eV again, and finally attains to 
1$\times$10$^8$ cm$^{-3}$ at $E_e$ = 100 eV.
The temperatures of C$_{60}^+$ and C$_{60}^-$, $T_+$ and $T_-$, are 
about 0.5 eV.
The plasma and floating potentials are almost 0 V which is equal to 
the potential of the grid and the thin annulus.
Therefore it can be said that the static potential-structures 
including sheaths are not formed in the pair-ion plasma 
because the ions have the same mass and temperature.
This property of the plasma generation in the energy range of 
$E_e$ $<$ 30 eV is quite different from the previous result 
for the case of the LaB$_6$ cathode \cite{Oohara3}.
The plasma density in $E_e$ $<$ 30 eV does not clearly depend on 
the magnetic field $B$, but strongly depends in $E_e$ $>$ 30 eV.
The direct contact of the fullerene vapor with the thermal cathode 
is supposed to be involved in the ionization process of fullerene, 
although the exact physical mechanism in $E_e$ $<$ 30 eV is not understood.
Since the density in Region (III) drastically decreases 
for $|V_{an}| \neq$ 0 V, the density modulation can be realized 
without using a grid immersed inside the plasma cross section, 
which disturbs the plasma condition.
Thus, longitudinal-electrostatic modes are excited in the pair-ion plasma, 
when the voltage of the annular exciter is temporally alternated 
($V_{exc} \neq 0$).

\section{Electrostatic Waves in Pair-Ion Plasma}

\hspace*{\parindent}
Some theoretical works have already been presented, 
which concern linear and nonlinear collective modes 
in non-relativistic electron-positron plasmas 
\cite{Stewart1,Iwamoto,Abdul,Stewart2,Zhao,Zank}.
A comprehensive two-fluid model has been developed for 
collective-mode analyses, based on which 
longitudinal (transverse) electrostatic (electromagnetic) modes 
have been studied.
The longitudinal collective modes are analogous to those 
in the ordinary electron-ion plasmas.
On the other hand, the transverse collective modes in the presence of 
a magnetic field are quite different from those in the ordinary plasmas, 
for instance, the whistler mode does not exist.
Here, electrostatic modes are focused in our pair-ion plasma, 
because the density and the temperature are relatively low and 
the induction current of the ions is very small, and electromagnetic modes 
relevant to the plasma can be neglected.

The twofluid equations in the absence of an applied $B$ field appropriate to 
the pair-ion plasma consist of the usual momentum and continuity equations 
for each species, supplemented by Poisson's equation.

\begin{equation}
mn_j ( \frac{\partial v_j}{\partial t} + (v_j\cdot\nabla)v_j) = 
-q_j n_j \nabla \phi - \gamma T_j \nabla n_j,
\end{equation}
\begin{equation}
\frac{\partial n_j}{\partial t} + \nabla \cdot (n_j v_j) = 0,
\end{equation}
\begin{equation}
\nabla ^2 \phi = -\frac{e}{\varepsilon_0} (n_+ - n_-).
\end{equation}
\vspace*{0mm}

\noindent
Where $m$, $n_j$, $v_j$, $q_j$, $T_j$, and $\phi$ denote the mass, 
the density, the fluid velocity, the charge, the temperature, 
and the potential, respectively. 
The subscript $j$ denotes positive or negative ions, $j$ = $+$ or $-$, 
$\gamma$ is the ratio of specific heats, and $\varepsilon _0$ is 
the permittivity of free space.
Linearizing about a homogeneous unbounded plasma 
(ion temperatures $T$ = $T_+$ = $T_-$), and defining 
the phase lag between the density fluctuations of positive and 
negative ions by $n_{+1}$ = $n_{-1}\exp(i\theta)$ for clarity, 
the coupled linear mode-equations are derived:

\begin{equation}
\omega^2 - c_s^2 k^2 - (1 - \exp (-i\theta)) \omega_p^2 = 0,
\end{equation}
\begin{equation}
\omega^2 - c_s^2 k^2 - (1 - \exp (i\theta)) \omega_p^2 = 0,
\end{equation}
\vspace*{0mm}

\noindent
where the acoustic speed $c_s^2$ = $\gamma T/m$ and the plasma frequency 
$\omega _p^2$ = e$^2$$n/\varepsilon_0 m$ are introduced.
The dispersion relations associated with Eqs.~(4) and (5) are simply given by

\begin{equation}
\omega^2 = c_s^2 k^2 \hspace{1cm} (\theta = 0),
\end{equation}
\begin{equation}
\omega^2 = c_s^2 k^2 + 2\omega_p^2 \hspace{1cm} (\theta = \pi).
\end{equation}
\vspace*{0mm}

\noindent
These modes are the ion acoustic wave (6) and the ion plasma wave (7).
In a nonzero applied $B$ field but for which a zero magnetic fluctuation, 
the dispersion relations \cite{Zank} are given by

\begin{equation}
(\omega^2 - \omega_p^2)(\omega^2 - c_s^2k^2 - \omega_h^2) + 
c_s^2 \omega_c^2 k^2 \cos ^2 \alpha = 0,
\end{equation}
\begin{equation}
\omega^2 (\omega^2 - c_s^2k^2 - \omega_c^2) + 
c_s^2 \omega_c^2 k^2 \cos ^2 \alpha = 0.
\end{equation}
\vspace*{0mm}

\noindent
These modes propagating along $B$-field lines ($\alpha$: propagation angle, 
$\alpha$ = 0) is 
quite same as that in the absence of $B$ field, 
except for the cyclotron oscillation 
$\omega _c$ and the upper hybrid oscillation 
$\omega_h = \sqrt{2\omega _p^2 + \omega _c^2}$.

The property of the wave propagation along the $B$-field lines 
is measured at $r$ = 0 cm and $z$ = 10$-$12 cm in Region (III) 
for $E_e$ = 100 eV by exciting the density fluctuation with 
the annular exciter, as mentioned above. 
The measured (dots) and the calculated (solid curves) dispersion relations 
are shown in Fig.~3, where the density modulation condition is 
$V_{an}$ = 0 V and $V_{exc}$ = 0.2 V, 
and the resultant amplitude of the density fluctuation $n_1$/$n_0$ is 
about 0.1.
The wave number (wave length) is obtained from the phase delay of 
the positive-ion density fluctuation measured 
at $z$ = 10, 11, and 12 cm.
The theoretical curves are calculated using Eqs.~(6) and (7) for 
$\gamma$ = 3 (one-dimensional compression), $T$ = 0.5 eV (isotropy), and 
$n_0$ = 1$\times$10$^7$ cm$^{-3}$.
Typical plasma parameters are as follows: the pair-plasma frequency is 
$(\sqrt{2}\omega_{p})/2\pi$ = 35 kHz, the cyclotron frequency 
($B$ = 0.3 T) is $\omega_{c}/2\pi$ = 6.4 kHz, and the acoustic speed is 
$c_s$ = 4.5$\times$10$^4$ cm/s.
There are three branches in the measured dispersion relation, 
$\omega/2\pi$ $<$ 8 kHz (the ion acoustic wave, IAW), 
8 $<$ $\omega/2\pi$ $<$ 32 kHz, and 
$\omega/2\pi$ $>$ 32 kHz (the ion plasma wave, IPW).
IPW measured fits to the calculated curve, and IAW measured 
in the relatively low-frequency band ($\omega/2\pi$ $<$ 3 kHz) also fits 
but deviates from it in the relatively high-frequency band 
(3 $<$ $\omega/2\pi$ $<$ 8 kHz).
It is worth while pointing out that a new mode is experimentally observed 
in 8 $<$ $\omega/2\pi$ $<$ 32 kHz for the first time, 
but we have not succeeded in deriving theoretically 
the dispersion relation of the new mode yet.
The new mode has the feature that the group velocity is negative 
but the phase velocity is positive, i.e., the mode is like a backward wave, 
and called the intermediate-frequency wave (IFW) here.

The typical temporal variations of the positive- and negative-ion densities 
and the potential are presented for $\omega /2\pi$ = 0.4, 2 kHz (IAW), 
20 kHz (IFW), and 35 kHz (IPW) in Figs.~4 (a), (b), (c), and (d), respectively.
$\tilde{I}_+$, $\tilde{I}_-$, and $\tilde{\phi}_f$ indicate 
the oscillating components of the positive and negative saturation currents 
(relative to the positive- and 
negative-ion densities), and the floating potential of the probe 
(the space potential) measured at $z$ = 10 cm, respectively.
The frequency spectrum of the phase lag 
$\theta(\tilde{I}_-) - \theta(\tilde{I}_+)$ between $\tilde{I}_+$ and 
$\tilde{I}_-$ is measured at $r$ = 0 cm and $z$ = 10 cm as shown in Fig.~5.
In Fig.~4 (a), the phase lag of the densities for IAW is close to zero 
and the amplitude of the potential oscillation is extremely small, 
which is consistent with the result derived from the theoretical dispersion 
relation (6).
Although the property of IAW measured in the very low-frequency range 
($\omega/2\pi$ $<$ 1 kHz) is the same as the theoretical one, 
the phase lag starts to increase in proportion to $\omega/2\pi$ 
(1 $<$ $\omega /2\pi$ $<$ 3 kHz) and attains to a constant value of about 
1.1$\pi$ (3 $<$ $\omega /2\pi$ $<$ 8 kHz), 
which has not been explained theoretically yet.
On the other hand the phase lags for IPW and IFW are $\pi$ and 1.03$\pi$ 
independently of $\omega/2\pi$, respectively, 
and their potential-oscillation amplitudes are large, as seen in 
Figs.~4 (c) and (d).

The modes propagating oblique $B$-field lines are observed in vicinity of the annular exciter.
The measured (dots) and the calculated (solid curves) dispersion relations are shown in Fig.~6, where the mode is measured at $r$ = 0 cm and $z$ = 2$-$4 cm for $E_e$ = 100 eV, $V_{an}$ = 0 V, and $V_{exc}$ = 0.2 V.
Here, the wave number $k_{||}$ denotes the parallel component of $k$ and 
the propagation angle is 0.48$\pi$ with respect to $B$-field lines.
The theoretical curves are calculated from Eqs.~(8) and (9) for 
$\alpha$ = 0.48$\pi$ and the plasma parameters are same as the case of propagating along $B$-field lines.
There are four branches and the ion cyclotron resonance appears 
in the measured dispersion relation.
A backward-wave like mode exists around the ion cyclotron frequency 
$\omega_c/2\pi$ (= 6.4 kHz) but not in the frequency range of between the ion 
acoustic wave and the ion plasma wave.
The phase lag of the modes propagating oblique $B$-field lines will be measured in our future works.

Some theoretical ideas for providing an explanation of the modes are recently 
proposed, and we introduce them here.
Dr. A. Hasegawa and Dr. P. K. Shukla suggest that the modes could be 
surface waves and they derive the dispersion relations of them 
in a planer model and a cylindrical model.
Dr. H. Schamel suggests the interpretation of the wave characteristics 
in term of periodic hole equilibria and associated kinetic modes, 
where the mode coupling between the ion plasma wave and the ion acoustic wave 
is caused by the particles trapped in the wave potential and 
IFW could appear.
Dr. F. Verheest suggests the oblique propagation of large amplitude 
electromagnetic solitons in pair plasmas.
Dr. J. Vranjes suggests that only two modes, the electrostatic Langmuir mode 
and the electromagnetic transverse plasma mode, could exist 
in the pair-ion plasma in the absence of $B$ field.
We will make it appear experimentally that these theoretical suggestions are 
well suited to this pair-ion plasma or not.

\section{Summary}

\hspace*{\parindent}
In summary, for the purpose of experimentally investigating electrostatic 
wave phenomena in a pair-ion plasma, 
the drastic improvement of the pair-ion plasma source consisting of only 
C$_{60}^+$ and C$_{60}^-$ with an equal mass is performed effectively 
using a magnetized hollow electron beam, and processes of 
electron-impact ionization, electron attachment, 
and magnetically-filtered separation.
The active excitation of density modulation by a thick annulus 
reveals the existence of three kinds of electrostatic modes propagating 
along $B$-field lines, an ion acoustic wave (IAW), an ion plasma wave (IPW), 
and an intermediate-frequency wave(IFW).
IAW and IPW are predicted in the twofluid theory for a unbounded plasma, 
while IFW is not predicted but only experimentally observed here. 
The phase lags between the density fluctuations of positive and 
negative ions are 0 for IAW in the low-frequency range, 
about $\pi$ for IFW, and $\pi$ for IPW.

\ \\
\noindent
{\bf Acknowledgement}
\ \\

 The authors would like to thank T.~Hirata and T.~Kaneko 
for their collaboration.
We are indebted to N.~Tomioka, M.~Kobayashi, and H.~Iwata 
for their support.
This work was supported by a Grant-in-Aid for Scientific Research 
from the Ministry of Education, Culture, Sports, Science, 
and Technology, Japan.
Finally, we thank some theoreticians for their suggestions.

\ \\

\clearpage

\begin{figure}
\begin{center}
\includegraphics[width=120mm]{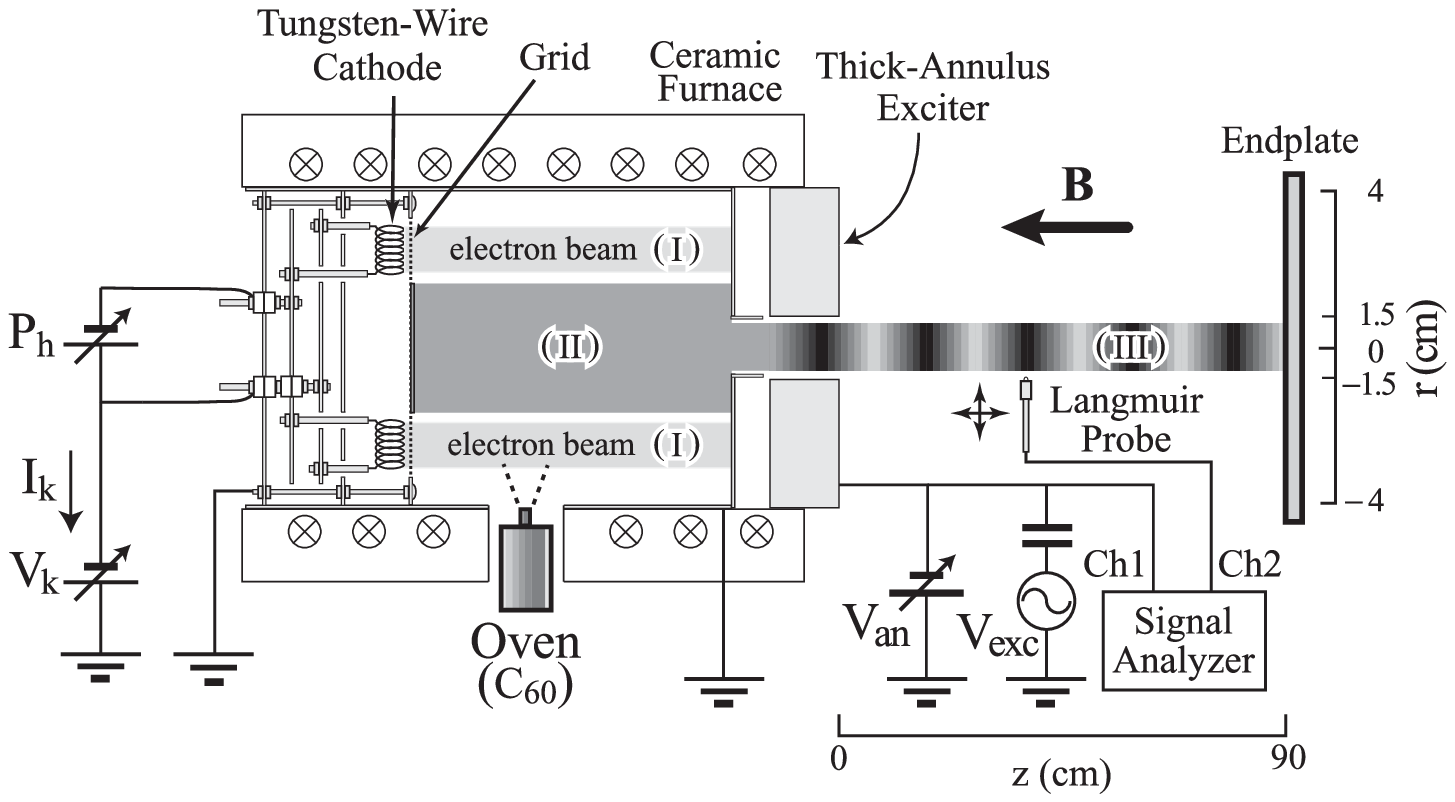}
\caption{Schematic drawing of the experimental setup.
Pure pair-ion plasma using fullerene (C$_{60}^+$, C$_{60}^-$) is 
generated by electron-impact ionization, electron attachment, 
and magnetic filtering.
Density modulation (longitudinal-electrostatic wave) is excited 
by thick annulus.} 
\end{center}
\end{figure}

\begin{figure}
\begin{center}
\includegraphics[width=60mm]{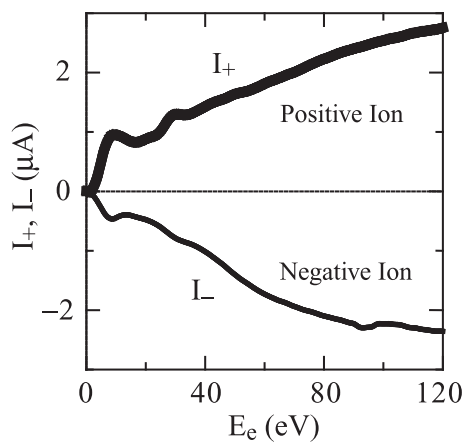}
\caption{Generation property of pair-ion plasma 
depending on electron-beam energy $E_e$ in Region (III).}
\end{center}
\end{figure}

\begin{figure}
\begin{center}
\includegraphics[width=60mm]{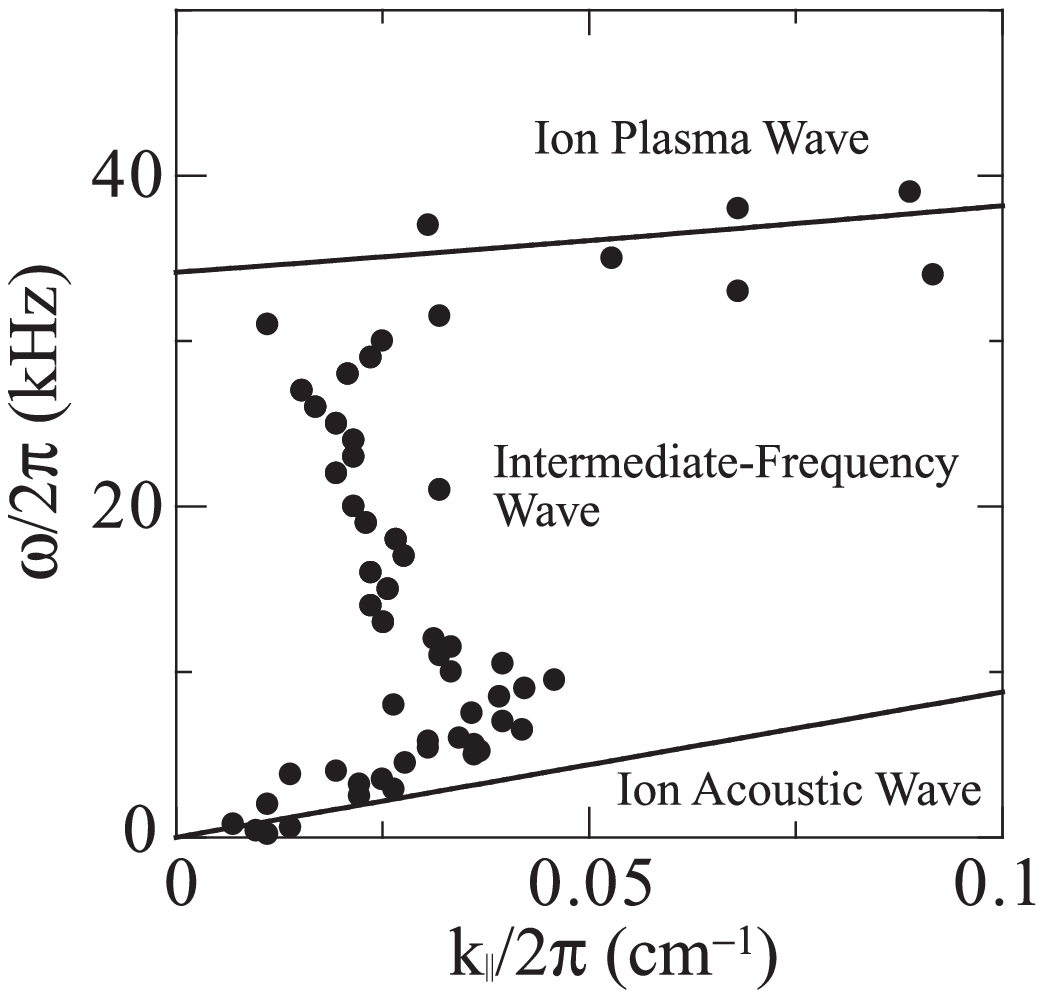}
\caption{Dispersion relations for electrostatic waves 
propagating along $B$-field lines. 
Dots denote experimental results.
Solid curves denote results calculated from Eqs. (6) and (7).}
\end{center}
\end{figure}

\begin{figure}
\begin{center}
\includegraphics[width=100mm]{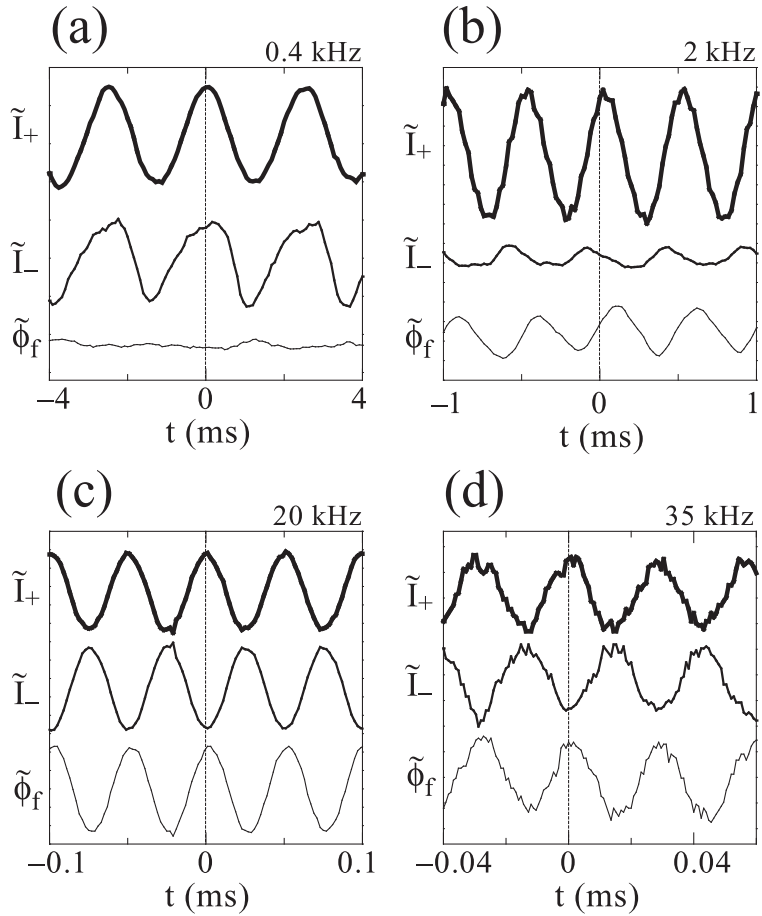}
\caption{Typical temporal variations of positive- and 
negative-ion densities and potential at 
$\omega/2\pi$ = (a) 0.4 kHz, (b) 2 kHz, (c) 20 kHz, and (d) 35 kHz.}
\end{center}
\end{figure}

\begin{figure}
\begin{center}
\includegraphics[width=60mm]{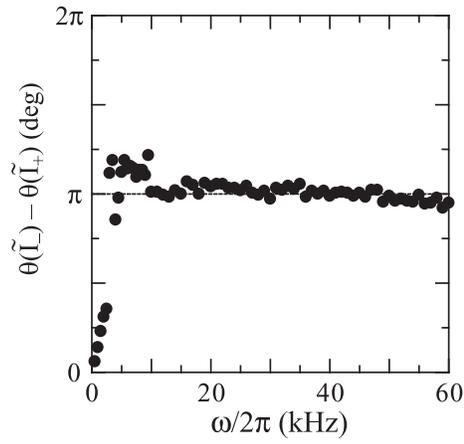}
\caption{Frequency spectrum of phase difference between negative and 
positive density fluctuations propagating along $B$-field lines measured around $z$ = 10 cm.}
\end{center}
\end{figure}

\begin{figure}
\begin{center}
\includegraphics[width=60mm]{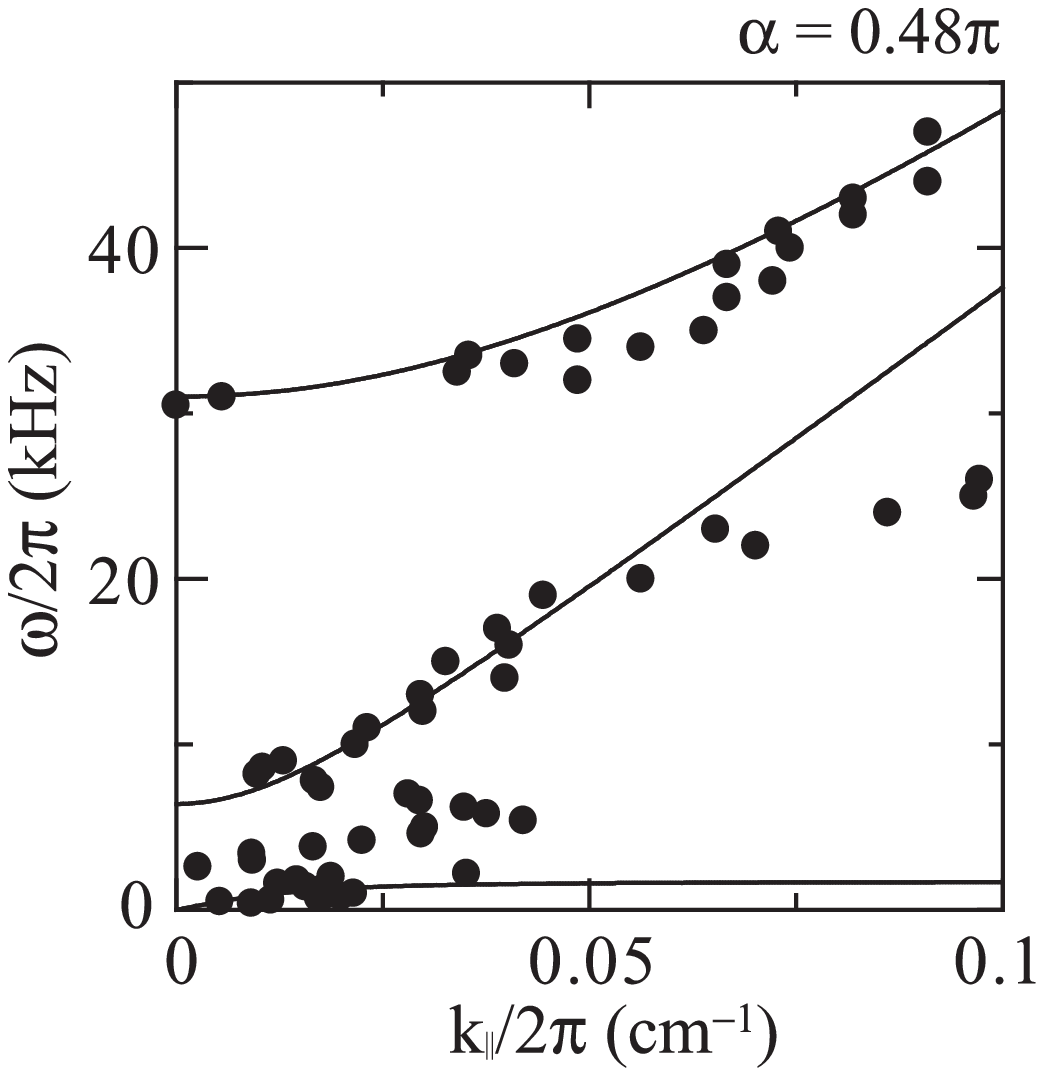}
\caption{Dispersion relations for electrostatic waves 
propagating oblique $B$-field lines. 
Dots denote experimantal results.
Solid curves denote results calculated from Eqs. (8) and (9).}
\end{center}
\end{figure}


\begin{thebibliography}{99}
\bibitem{Gibson}
G. Gibson, W. C. Jordan, and E. J. Lauer,
Phys. Rev. Lett. {\bf 5}, 141 (1960).

\bibitem{Surko1}
C. M. Surko, M. Leventhal, and A. Passner,
Phys. Rev. Lett. {\bf 62}, 901 (1989).

\bibitem{Surko2}
C. M. Surko and T. J. Murphy,
Phys. Fluids B {\bf 2}, 1372 (1990).

\bibitem{Greaves}
R. G. Greaves, M. D. Tinkle, and C. M. Surko,
Phys. Plasmas {\bf 1}, 1439 (1994).

\bibitem{Boehmer}
H. Boehmer, M. Adams, and N. Rynn,
Phys. Plasmas {\bf 2}, 4369 (1995).

\bibitem{Liang}
E. P. Liang, S. C. Wilks, and M. Tabak, Phys. Rev. Lett. 
{\bf 81}, 4887 (1998).

\bibitem{Amoretti}
M. Amoretti et al, Phys. Rev. Lett. 
{\bf 91}, 055001 (2003).

\bibitem{Stewart1}
G. A. Stewart and E. W. Laing, 
J. Plasma Phys. {\bf 47}, 295 (1992).

\bibitem{Iwamoto}
N. Iwamoto, Phys. Rev. E {\bf 47}, 604 (1993).

\bibitem{Abdul}
S. Y. Abdul-Rassak and E. W. Laing, 
J. Plasma Phys. {\bf 50}, 125 (1993).

\bibitem{Stewart2}
G. A. Stewart, 
J. Plasma Phys. {\bf 50}, 521 (1993).

\bibitem{Zhao}
J. Zhao, K. I. Nishikawa, J. I. Sakai, and T. Neubert, 
Phys. Plasmas {\bf 1}, 103 (1994).

\bibitem{Zank}
G. P. Zank and R. G. Greaves, Phys. Rev. E {\bf 51}, 6079 (1995).

\bibitem{Dubin}
D. H. E. Dubin, Phys. Rev. Lett. 
{\bf 92}, 195002 (2004).

\bibitem{Schermann}
J. P. Schermann and F. G. Major, 
Appl. Phys. {\bf 16}, 225 (1978).

\bibitem{NSato}
N. Sato, T. Mieno, T. Hirata, Y. Yagi, R. Hatakeyama, and S. Iizuka,
Phys. Plasmas {\bf 1}, 3480 (1994).

\bibitem{Oohara1}
W. Oohara, R. Hatakeyama, and S. Ishiguro,
Plasma Phys. Control. Fusion {\bf 44}, 1299 (2002).

\bibitem{Oohara2}
W. Oohara, R. Hatakeyama, and S. Ishiguro, 
Phy. Rev. E {\bf 68}, 066407 (2003).

\bibitem{Jaffke}
T. Jaffke, E. Illenberger, M. Lezius, S. Matejcik, D. Smith, and 
T. D. M$\ddot{\rm a}$rk,
Chem. Phys. Lett. {\bf 226}, 213 (1994).

\bibitem{Huang}
J. Huang, H. S. Carman, Jr., and R. N. Compton,
J. Phys. Chem. {\bf 99}, 1719 (1995).

\bibitem{Volpel}
R. V$\ddot{\rm o}$lpel, G. Hofmann, M. Steidl, M. Stenke, M. Schlapp, 
R. Trassl, and E. Salzborn,
Phys. Rev. Lett. {\bf 71}, 3439 (1993).

\bibitem{Matt}
S. Matt, B. D$\ddot{\rm u}$nser, M. Lezius, H. Deutsch, K. Becker, 
A. Stamatovic, P. Scheier, and T. D. M$\ddot{\rm a}$rk,
J. Chem. Phys. {\bf 105}, 1880 (1996).

\bibitem{Vostrikov}
A. A. Vostrikov, D. Yu. Dubnov, and A. A. Agarkov, 
High Temperature {\bf 39}, 22 (2001).

\bibitem{Oohara3}
W. Oohara and R. Hatakeyama, 
Phys. Rev. Lett. {\bf 91}, 205005 (2003).

\bibitem{Sheehan}
D. P. Sheehan and N. Rynn, 
Rev. Sci. Instrum. {\bf 59}, 1369 (1988).

\end{thebibliography}
\end{document}